\documentclass[10pt,conference]{IEEEtran}
\IEEEoverridecommandlockouts
\usepackage{amsmath,amsfonts}
\usepackage{algorithmic}
\usepackage{algorithm}
\usepackage{array}
\usepackage[caption=false,font=normalsize,labelfont=sf,textfont=sf]{subfig}
\usepackage{textcomp}
\usepackage{stfloats}
\usepackage{url}
\usepackage{verbatim}
\usepackage{graphicx}
\usepackage{cite}
\usepackage{listings, xcolor}
\definecolor{verylightgray}{rgb}{.97,.97,.97}
\lstdefinelanguage{Solidity}{
  keywords=[1]{anonymous, assembly, assert, balance, break, call, callcode, case, catch, class, constant, continue, constructor, contract, debugger, default, delegatecall, delete, do, else, emit, event, experimental, export, external, false, finally, for, function, gas, if, implements, import, in, indexed, instanceof, interface, internal, is, length, library, log0, log1, log2, log3, log4, memory, modifier, new, payable, pragma, private, protected, public, pure, push, require, return, returns, revert, self-destruct, send, solidity, storage, struct, suicide, super, switch, then, this, throw, transfer, true, try, typeof, using, value, view, while, with, addmod, ecrecover, keccak256, mulmod, ripemd160, sha256, sha3}, 
  keywordstyle=[1]\color{blue}\bfseries,
  keywords=[2]{address, bool, byte, bytes, bytes1, bytes2, bytes3, bytes4, bytes5, bytes6, bytes7, bytes8, bytes9, bytes10, bytes11, bytes12, bytes13, bytes14, bytes15, bytes16, bytes17, bytes18, bytes19, bytes20, bytes21, bytes22, bytes23, bytes24, bytes25, bytes26, bytes27, bytes28, bytes29, bytes30, bytes31, bytes32, enum, int, int8, int16, int24, int32, int40, int48, int56, int64, int72, int80, int88, int96, int104, int112, int120, int128, int136, int144, int152, int160, int168, int176, int184, int192, int200, int208, int216, int224, int232, int240, int248, int256, mapping, string, uint, uint8, uint16, uint24, uint32, uint40, uint48, uint56, uint64, uint72, uint80, uint88, uint96, uint104, uint112, uint120, uint128, uint136, uint144, uint152, uint160, uint168, uint176, uint184, uint192, uint200, uint208, uint216, uint224, uint232, uint240, uint248, uint256, var, void, ether, finney, szabo, wei, days, hours, minutes, seconds, weeks, years},  
  keywordstyle=[2]\color{teal}\bfseries,
  keywords=[3]{block, blockhash, coinbase, difficulty, gaslimit, number, timestamp, msg, data, gas, sender, sig, value, now, tx, gasprice, origin},  
  keywordstyle=[3]\color{violet}\bfseries,
  identifierstyle=\color{black},
  sensitive=false,
  comment=[l]{//},
  morecomment=[s]{/*}{*/},
  commentstyle=\color{gray}\ttfamily,
  stringstyle=\color{red}\ttfamily,
  morestring=[b]',
  morestring=[b]"
}
\usepackage{breakurl}
\usepackage{graphicx}
\usepackage{hyperref}

\usepackage{soul}

\begin{document}

\title{SSR: Safeguarding Staking Rewards by Defining and Detecting Logical Defects in DeFi Staking}




\author{\IEEEauthorblockN{Zewei Lin \IEEEauthorrefmark{2}\IEEEauthorrefmark{4},
Jiachi Chen\IEEEauthorrefmark{2}, 
Jingwen Zhang\IEEEauthorrefmark{2}\IEEEauthorrefmark{4},
Zexu Wang\IEEEauthorrefmark{2}\IEEEauthorrefmark{4},
Yuming Feng\IEEEauthorrefmark{4},
Weizhe Zhang\IEEEauthorrefmark{3}\IEEEauthorrefmark{4}, and
Zibin Zheng\thanks{* Zibin Zheng is the corresponding author.}\IEEEauthorrefmark{2}\IEEEauthorrefmark{1}
}
\IEEEauthorblockA{\IEEEauthorrefmark{2}Sun Yat-sen University, \href{mailto:zhangjw273@mail2.sysu.edu.cn,liwei378@mail2.sysu.edu.cn,ningkw@mail2.sysu.edu.cn,linzw3@mail2.sysu.edu.cn,yaozt@mail2.sysu.edu.cn}{\{linzw3,  zhangjw273, wangzx97\}@mail2.sysu.edu.cn}, \href{mailto:nanyh@mail.sysu.edu.cn,zhzibin@mail.sysu.edu.cn}{\{chenjch86, zhzibin\}@mail.sysu.edu.cn}\\
\IEEEauthorrefmark{3}Harbin Institute of Technology, \href{mailto:wzzhang@hit.edu.cn}{wzzhang@hit.edu.cn}, \\
\IEEEauthorrefmark{4}Peng Cheng Laboratory,\href{mailto:fengym@pcl.ac.cn}{ fengym@pcl.ac.cn}
}}


\maketitle

\begin{abstract}
Decentralized Finance (DeFi) staking is one of the most prominent applications within the DeFi ecosystem, where DeFi projects enable users to stake tokens on the platform and reward participants with additional tokens.
However, logical defects in DeFi staking could enable attackers to claim unwarranted rewards by manipulating reward amounts, repeatedly claiming rewards, or engaging in other malicious actions.
To mitigate these threats, we conducted the first study focused on defining and detecting logical defects in DeFi staking. Through the analysis of 64 security incidents and 144 audit reports, we identified six distinct types of logical defects, each accompanied by detailed descriptions and code examples.
Building on this empirical research, we developed SSR (\textbf{S}afeguarding \textbf{S}taking \textbf{R}eward), a static analysis tool designed to detect logical defects in DeFi staking contracts. SSR utilizes a large language model (LLM) to extract fundamental information about staking logic and constructs a DeFi staking model. It then identifies logical defects by analyzing the model and the associated semantic features.
We constructed a ground truth dataset based on known security incidents and audit reports to evaluate the effectiveness of SSR. The results indicate that SSR achieves an overall precision of 92.31\%, a recall of 87.92\%, and an F1-score of 88.85\%.
Additionally, to assess the prevalence of logical defects in real-world smart contracts, we compiled a large-scale dataset of 15,992 DeFi staking contracts. SSR detected that 3,557 (22.24\%) of these contracts contained at least one logical defect.

\end{abstract}


\section{Introduction}
In recent years, Decentralized Finance (DeFi) has experienced rapid growth, with DeFi staking emerging as one of its most prominent applications. DeFi staking refers to the practice whereby DeFi projects enable users to stake tokens on the platform in order to increase the total value locked (TVL)~\cite{tvl} and reward participants with additional tokens~\cite{tokenomics}. This mechanism serves two primary purposes: for project developers, it helps increase the project’s token TVL, thereby supporting token value and attracting investment; for users, it offers an opportunity to earn additional rewards.

Logical defects in DeFi staking can allow attackers to claim unwarranted rewards. 
Logical defects arise from errors or incomplete logic in the contract, while non-logical defects result from external environmental factors or improper operations~\cite{liu2025exploring,liao2022large}.
Logical defects in DeFi staking can be exploited to manipulate staking reward amounts, repeatedly claim rewards, and perform other malicious actions~\cite{22reviewStaking, lendingStakingVul}. 
To date, logical defects in DeFi staking contracts have led to substantial financial losses.
In October 2024, the DeFi staking project OTSea Staking suffered a loss of \$26,000 due to a logical defect that allowed manipulation of the variable used in calculating staking rewards~\cite{otseastakingIncident}.

To the best of our knowledge, no prior research has focused on the logical defects in DeFi staking. 
Existing studies on DeFi staking primarily focus on its tokenomics, which combines aspects of tokens and economics~\cite{wikiTokenomics}, including token transactions and distribution~\cite{carre2024liquid, cong2025tokenomics}. 
Meanwhile, current methods for detecting contract defects are unable to detect logical defects in DeFi staking, as they cannot identify logic related to staking, such as the formulas used to calculate rewards~\cite{chen2021defectchecker, sun2024gptscan}.
These logics are defined according to unique requirements, making them difficult to detect using predefined rules.
Furthermore, security analyses of DeFi staking incidents often fail to identify the root causes. For instance, while some analyses note manipulated staking rewards, they neglect to examine the code leading to the manipulation~\cite{autosharkExample}.

To address this gap, an empirical study on logical defects in DeFi staking is conducted. 64 security incidents and 144 audit reports related to DeFi staking are collected and analyzed. 
Two researchers manually examined these incidents and reports using the open card sorting method. Based on the root causes identified, six types of logical defects in DeFi staking are defined: \textit{Staking Logical Variables Manipulation (SVM)},
\textit{Rewards without TimeDelay (RT)},
\textit{Single Liquidity Pool Reliance (SLR)},
\textit{Omission in Status Update (OSU)},
\textit{Unsafe Verification (UV)},
and \textit{Unauthorized Staking Asset Access (UAA)}.

Based on the empirical study, we developed a detection tool for logical defects in DeFi staking, named SSR (\textbf{S}afeguarding \textbf{S}taking \textbf{R}eward). 
Given the source code of a DeFi staking contract, SSR first employs a large language model (LLM) to extract basic staking-related information. 
It then constructs a comprehensive model of the DeFi staking logic based on the extracted data and the contract’s control flow graph, and identifies relevant semantic features. 
Finally, leveraging the DeFi staking model and the extracted semantic features, SSR detects six types of logical defects using predefined rules.

To evaluate the effectiveness of SSR in detecting logical defects in DeFi staking, we constructed a ground truth dataset consisting of 40 DeFi staking smart contracts, derived from the security incidents and audit reports used in the empirical study.
On this dataset, SSR achieved an overall precision of 92.31\%, a recall of 87.92\%, and an F1-score of 88.85\%. 
To further assess the prevalence of logical defects in real-world DeFi staking contracts, we developed a large-scale dataset comprising 15,992 open-source DeFi staking smart contracts.
Our analysis revealed that 3,557 (22.24\%) of these contracts contained at least one logical defect. 
We assessed SSR’s performance on the large-scale dataset through random sampling and manual verification. The experimental results indicated that SSR achieved an overall precision of 89.41\%.

The primary contributions of this work are as follows:
\begin{itemize}
\item{We conducted the first study on logical defects in DeFi staking. By analyzing 64 security incidents and 144 audit reports, we identified and classified six types of logical defects, thereby expanding the existing classification~\cite{sun2024gptscan}. We provided detailed descriptions and code examples of these defects to inform future security solutions.}
\item{We proposed SSR, a tool for detecting logical defects in DeFi staking. It extracts fundamental information about DeFi staking using an LLM and constructs a DeFi staking model for logical defect detection. On the ground truth dataset, it achieved an overall precision of 92.31\%, recall of 87.92\%, and an F1-score of 88.85\%.}
\item{We collected a large-scale dataset of 15,992 real-world DeFi staking smart contracts to further assess the prevalence of logical defects. SSR revealed that 3,557 (22.24\%) of these contracts contain at least one logical defect, achieving an overall precision of 89.41\%.}
\item{To support future research, we published the source code of SSR, along with all analysis results and datasets at https://github.com/Zer0zw/SSR.}
\end{itemize}

\section{Background}
\subsection{Smart Contract and Decentralized Finance (DeFi)}
A smart contract is a program deployed on a blockchain~\cite{nakamoto2008bitcoin} that autonomously executes actions based on predefined conditions, without the need for intermediaries~\cite{zheng2020overview, zheng2018blockchain}. 
Due to the inherent characteristics of smart contracts, defects within them can lead to significant economic losses.


Decentralized Finance (DeFi) refers to financial applications built on blockchains, such as Ethereum~\cite{etherscan}, using smart contracts. Its aim is to replicate the functions of traditional financial institutions, such as banks, exchanges, and lending platforms, in a decentralized manner~\cite{werner2022sok}. Among DeFi applications, staking is one of the most prominent

\subsection{DeFi Staking}
DeFi staking refers to the practice in which DeFi projects allow users to stake tokens on the platform to increase the total value locked (TVL) and reward participants with additional tokens~\cite{22reviewStaking}.
Tokenomics in DeFi staking refers to the design of blockchain-based economic systems, including the token distribution, utility, and rewards that influence user behavior~\cite{tokenomics}.

In DeFi staking smart contracts, three core functionalities are typically implemented: staking, claiming rewards, and unstaking~\cite{9defistakingplat}.
Staking involves users depositing their cryptocurrency assets into a platform’s staking pool or smart contract.
Claiming rewards is the process through which users receive incentives, usually in the form of additional tokens, based on the proportion of their staked assets contributed to the DeFi project.
Unstaking enables users to withdraw their deposited assets after completing the required lock-up period.

\section{Logical Defects in DeFi Staking}~\label{sec:defects}
To gain deeper insights into the logical defects of DeFi staking and to design effective detection methods, we conducted an empirical study.
Based on the findings, we classified these defects into six distinct categories.
This section presents the classification criteria and methodology, followed by precise definitions and illustrative examples for each category of logical defect in DeFi staking.

\subsection{Data Collection}
\subsubsection{DeFi Staking Security Incidents}
To identify the logical defects affecting DeFi staking contracts, we collected and analyzed security incidents involving DeFi staking projects.
We searched for relevant security incidents using the keywords “stake” and “staking” on blockchain security platforms, including DeFiHackLab~\cite{defihacklab}, SlowMist~\cite{slowmist}, Blocksec~\cite{blocksec}, Peckshield~\cite{peckshield}, and Certik~\cite{certik}. As a result, we identified 127 security incidents associated with DeFi staking projects.

\subsubsection{Smart Contract Audit Reports}
To gain a more comprehensive understanding of the logical defects in DeFi staking, we also analyzed audit reports to evaluate which defects had been mitigated through audits. 
We collected audit reports from two sources: the DAppScan~\cite{zheng2024dappscan} dataset and smart contract audit platforms.
DAppScan is a large-scale dataset of weaknesses from Decentralized Applications (DApps)~\cite{dapps}, accompanied by audit reports from various security teams.
We searched for audit reports within the DAppScan dataset using keywords. 
In addition, we collected audit reports from Cyberscope~\cite{cyberscope}, CertiK~\cite{certik}, and other smart contract auditing platforms. 
In total, we collected 201 relevant audit reports.

\subsection{Data Analysis}
\subsubsection{Manually Filtering}
To conduct a more accurate analysis of the logical defects in DeFi staking, data unrelated to these defects is filtered out. Two researchers manually examined and filtered the collected security incidents and smart contract audit reports. 
For security incidents, we analyzed the root cause of the incident, excluding cases where the targeted contract was not a DeFi staking contract or where the incident’s cause was unrelated to logical defects (e.g., private key leakage, Ponzi schemes). 
For audit reports, we excluded those that did not focus on DeFi staking contracts. 
After this filtering process, we identified 64 security incidents and 144 audit reports.

\subsubsection{Open Card Sorting}
To ensure accurate and comprehensive classification, we employed the open card sorting method~\cite{spencer2009card}, which is commonly used in empirical research~\cite{chen2020defining, zhang2024demystifying}, to categorize the collected incidents and reports.
First, cards were created for each incident and report, each card containing the project name, a description of the defect, and the root cause (i.e., the code snippets that exhibited the defect). Then, two researchers, each with more than three years of experience in smart contract research, manually analyzed and classified these cards. The open card sorting process consisted of two rounds.


In the first round, 40\% of the cards were randomly selected and jointly analyzed by both researchers to establish an initial classification. 
During the analysis, the researchers initially analyzed the defective code snippets guided by the descriptions of the defects. 
In the case of attack events, they also examined the specific attack processes. Subsequently, the cards were categorized based on the root cause of each defect. 
For instance, in cases of staking reward manipulation, the researchers investigated the precise reason for the manipulation, such as whether it involved altering a key variable in the reward calculation or relying on a single liquidity pool. 
This approach emphasized identifying the root cause of each defect, rather than broadly grouping cases under general labels like ``staking reward manipulation''. 
Cards for which the root cause could not be identified were excluded from the classification.

In the second round, the remaining 60\% of the cards were analyzed and classified independently by each researcher. 
The steps for analyzing the cards in this round were identical to those used in the first round.
Discrepancies in classification were subsequently discussed collaboratively to finalize the categorization of logical defects in DeFi staking.

\vspace{-0.3cm}
\begin{figure}[h]
    \centering
    \setlength{\abovecaptionskip}{0.05cm}
    \includegraphics[width=\linewidth]{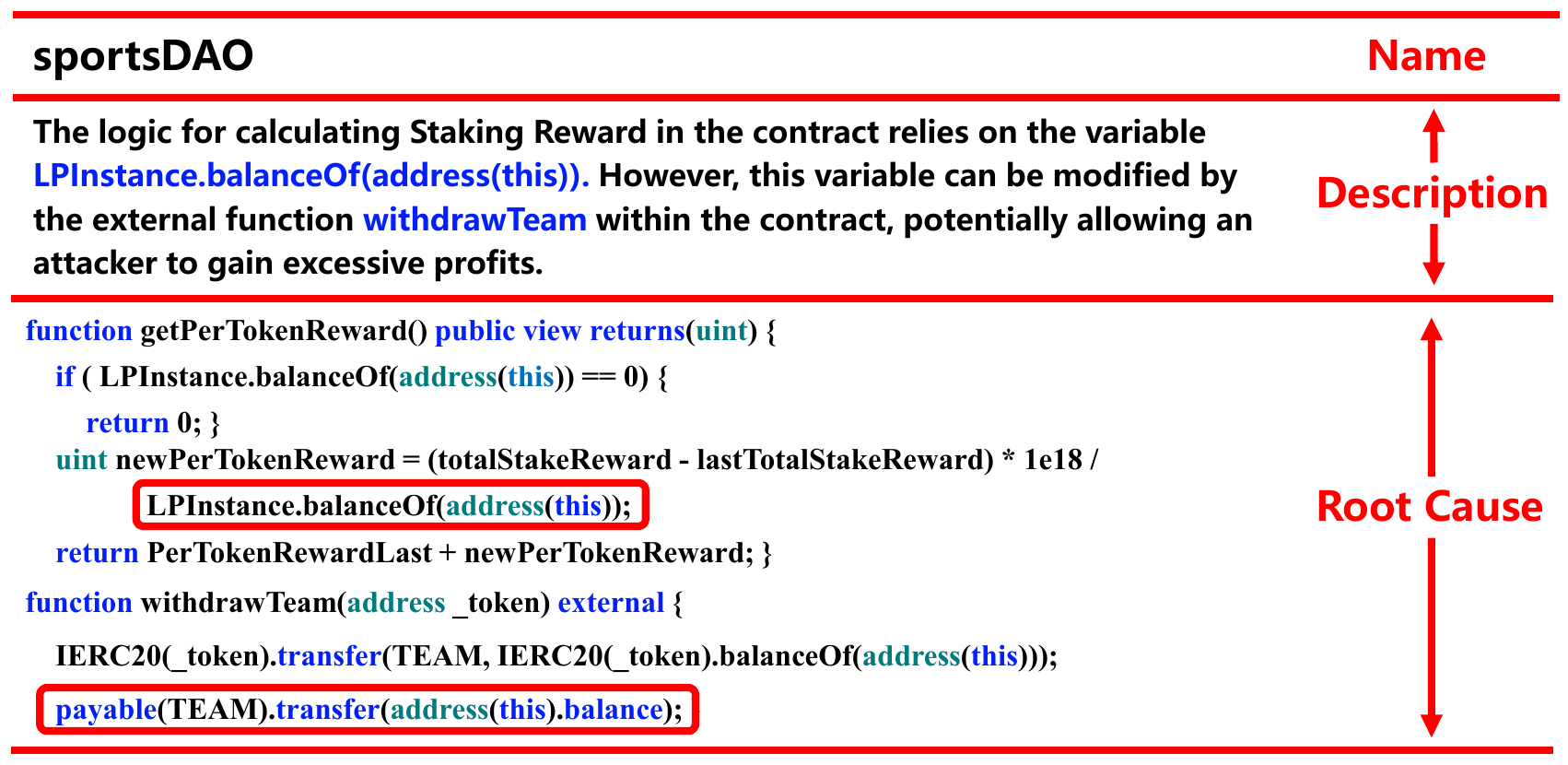}
    \caption{Example of a card of DeFi staking security incident}
    \label{fig:card}
\end{figure}
\vspace{-0.3cm}

Figure~\ref{fig:card} illustrates a card example generated from a DeFi staking security incident. 
In this case, the targeted project is sportsDAO~\cite{bscscanSportsdao}, where staking rewards are calculated based on the token balance within the contract. 
However, the \textit{withdrawTeam} function allows any user to transfer all staking tokens to the contract owner, reducing the token balance to zero. 
As the token balance decreased significantly, the staking rewards available to the attacker increased accordingly.
Based on root cause analysis, this issue is classified as a case of \textit{Staking Logical Variables Manipulation}.

\subsection{Definition of Logical Defects in DeFi Staking}
Based on open card sorting, we identified six distinct types of logical defects in decentralized finance (DeFi) staking. These defects may allow attackers to exploit reward calculations, repeatedly claim rewards, or steal others’ staked assets and rewards. 
Table~\ref{tab:defects_def} provides concise definitions of these six defect types. 
Below, we provide detailed explanations and code examples for each defect.

It is important to note that this paper focuses exclusively on logical defects in DeFi staking and does not address non-logical defects, such as reentrancy. Furthermore, if a DeFi staking contract contains multiple logical defects, each defect can independently lead to a security vulnerability. While multiple defects may coexist within a single contract, our analysis of 64 security incidents revealed that each incident was ultimately caused by a specific defect.

\vspace{-0.4cm}
\begin{table}[h] 
    \centering
    \setlength{\abovecaptionskip}{0.05cm}
        \caption{Definition of Logical Defects in DeFi Staking.}
        \resizebox{\linewidth}{!}{
        \begin{tabular}{p{3.1cm}|p{5.2cm}}
            \hline
            \textbf{Logical Defect} & \textbf{Definition} \\
            \hline
            \textit{Staking Logical Variables Manipulation (SVM)} & The values of certain critical variables related to DeFi staking logic can be arbitrarily modified by external functions. \\
            \hline
            \textit{Rewards without Timedelay (RT)} & Staking rewards are calculated without considering the staking duration, and no temporal restriction is imposed on claiming rewards. \\
            \hline
            \textit{Single Liquidity Pool Reliance (SLR)} & Staking rewards and minted staking tokens depend solely on a single liquidity pool.\\
            \hline
            \textit{Omission in Status Update (OSU)} & The relevant staking state is not updated in a timely manner. \\
            \hline
            \textit{Unsafe Verification (UV)} & A lack of proper validation checks or insufficient validation conditions when claiming rewards or unstaking tokens. \\
            \hline
            \textit{Unauthorized Staking Asset Access (UAA)} & The existence of functions that allow unauthorized access to users’ staking asset. \\
            \hline

        \end{tabular}}
    \label{tab:defects_def}
\end{table}
\vspace{-0.3cm}

\noindent\textbf{(1) Staking Logical Variables Manipulation (SVM):} 
This defect arises because certain critical variables in the DeFi staking system can be arbitrarily modified. 
Depending on the role of these variables, such modifications can lead to various consequences. 
For example: manipulating the amount of staking rewards; enabling or disabling staking or reward-claiming functions; and altering fee rates.
Such manipulations can be exploited by attackers to unjustly obtain rewards or alter the operational states.
It should be noted that this defect does not include normal user staking or unstaking operations.

\vspace{-0.4cm}
\begin{figure}[h]
\setlength{\abovecaptionskip}{0.05cm}
\begin{lstlisting}[language=Solidity,mathescape]
mapping(address => uint) public userStakeAmount;
uint256 public rewardRate
function claimReward() public {
    uint _reward = userStakeAmount[_msgSender()].mul(rewardRate);
    require(_reward > 0, "Stake Reward is 0");
    _standardTransfer(address(this), _msgSender(), _reward); }
function setRewardRate(uint _newRewardRate) public {
		rewardRate = _newRewardRate; }
\end{lstlisting}
\caption{An example of Staking Logical Variables Manipulation defect.}
\label{fig:defect_svm}
\end{figure}
\vspace{-0.3cm}

\textbf{Example:} 
Figure~\ref{fig:defect_svm} presents a simplified example of the \textit{Staking Logical Variables Manipulation} defect. 
In this example, users claim rewards by invoking the $claimReward$ function (lines 3–6), where the reward amount is calculated as the product of the staked token amount and $rewardRate$ (line 4). However, the external function $setRewardRate$ allows arbitrary modification of $rewardRate$ (line 8).
An attacker can exploit this defect by adjusting the reward rate before claiming rewards, thereby obtaining unjustified profits.

\noindent\textbf{(2) Rewards without Timedelay (RT):} 
This defect arises because the calculation of DeFi staking rewards does not consider the staking duration, and there are no time restrictions on claiming them.
Since the rewards are not tied to the staking duration, an attacker can repeatedly execute the sequence of stake, claim rewards, and unstake within a single block.
Each iteration yields a reward, allowing the attacker to unjustly accumulate rewards multiple times.

\vspace{-0.4cm}
\begin{figure}[h]
\setlength{\abovecaptionskip}{0.05cm}
\begin{lstlisting}[language=Solidity,mathescape]
mapping(address => uint) public lastClaimTime;
function claimReward() public {
		uint stakeDuration = block.timestamp - lastClaimTime[_msgSender()]
    uint _reward = stakeAmount[_msgSender()].mul(stakeDuration).mul(1e18).div(totalSupply());
    lastClaimTime[_msgSender()] = block.timestamp
    _standardTransfer(address(this), _msgSender(), _reward); }
\end{lstlisting}
\caption{An example of Rewards with Timedelay.}
\label{fig:defect_rt}
\end{figure}
\vspace{-0.3cm}

\textbf{Example:}
Figure~\ref{fig:defect_rt} illustrates a simplified example of reward calculation that considers staking duration.
In this case, the staking rewards are calculated based on both the user’s stake amount and the duration of the staking period (line 4).
By incorporating stake duration into the reward calculation, rewards remain zero in the block where tokens are staked, thereby preventing exploitation through repeated cycles of staking, claiming rewards, and unstaking. 
Additionally, by updating the timestamp of the last reward claim (line 5), any subsequent claims within the same block yield zero rewards, thereby mitigating exploitation through multiple invocations of the $claimReward$ function.

\noindent\textbf{(3) Single Liquidity Pool Reliance (SLR):} 
This defect arises from the calculation of staking rewards or the amount of staking tokens minted based on a single liquidity pool.
An attacker can exploit this defect by initiating a flash loan attack to manipulate the token ratio within the liquidity pool and subsequently inflate the calculated rewards.

\vspace{-0.4cm}
\begin{figure}[h]
\setlength{\abovecaptionskip}{0.05cm}
\begin{lstlisting}[language=Solidity,mathescape]
IPancakePair private _lp;
function rewardTokenPriceInToken() public view returns (uint256) {
    (uint256 lpReserve0, uint256 lpReserve1, uint256 lpTimestamp) = _lp.getReserves();
    return (_tokenLpIndex == 0 ? lpReserve0 : lpReserve1).mul(1e18).div(_tokenLpIndex == 0 ? lpReserve1 : lpReserve0); }
\end{lstlisting}
\caption{An example of Single Liquidity Pool Reliance defect.}
\label{fig:defect_slr}
\end{figure}
\vspace{-0.3cm}

\textbf{Example:}
Figure~\ref{fig:defect_slr} presents a simplified example of the \textit{Single Liquidity Pool Reliance} defect. In the function $rewardTokenPriceInToken$ (lines 2–4), the staking reward is calculated based on the token ratio in the liquidity pool, obtained via $\_lp.getReserves()$ (line 3). 
Since the rewards rely solely on a single liquidity pool, an attacker can manipulate the token ratio within it and thereby inflate the rewards.

\noindent\textbf{(4) Omission in Status Update (OSU):} 
This defect refers to the failure to promptly update DeFi staking-related status. Specifically, it involves variables such as staking duration and account balance not being updated immediately after actions such as staking, claiming rewards, or unstaking.

\textbf{Example:}
Figure~\ref{fig:defect_osu} presents a simplified example of the \textit{Omission in Status Update} defect. In this case, the $unStake$ function enables users to unstake tokens and transfers a specified amount of staked tokens to the user (line 5). However, the user stake amount has not been updated (line 4). As a result, an attacker can repeatedly invoke the $unStake$ function to unjustly receive staked tokens multiple times.

\vspace{-0.4cm}
\begin{figure}[h]
\setlength{\abovecaptionskip}{0.05cm}
\begin{lstlisting}[language=Solidity,mathescape]
function unStake(uint _amount) public {
		uint _userBalances=userStakeAmount[msg.sender];
		require(_userBalances >= _amount, "User Stake Amount not enough");
		// userStakeAmount[msg.sender] = _userBalances.sub(_amount);
    _standardTransfer(address(this), msg.sender, _amount); }
\end{lstlisting}
\caption{An example of Omission in Status Update defect.}
\label{fig:defect_osu}
\end{figure}
\vspace{-0.3cm}

\noindent\textbf{(5) Unsafe Verification (UV):} 
This defect refers to the lack of necessary validation conditions required for claiming rewards or unstaking tokens.
For instance, this may occur when the contract fails to verify the staking status before executing an unstake or reward claim operation.

\vspace{-0.4cm}
\begin{figure}[h]
\setlength{\abovecaptionskip}{0.05cm}
\begin{lstlisting}[language=Solidity,mathescape]
function claimReward(uint256 stakeIndex) external {
    require(stakeIndex < stakers[msg.sender].length, "Invalid stake index");
    // require(stakers[msg.sender][stakeIndex].isUnstake, "Already unstaked");
    reward = calculateReward(msg.sender,stakeIndex);
    require(reward > 0, "No rewards to claim"); 
    stakeInfo.rewardsClaimed = stakeInfo.rewardsClaimed.add(reward);
    rewardToken.transfer(msg.sender, reward);  }
\end{lstlisting}
\caption{An example of Unsafe Verification defect.}
\label{fig:defect_uv}
\end{figure}
\vspace{-0.3cm}

\textbf{Example:}
Figure~\ref{fig:defect_uv} illustrates a simplified example of the \textit{Unsafe Verification} defect. 
In this example, users can claim rewards for the staked NFTs using the function $claimReward$. 
However, the validation to verify whether the NFT has already been unstaked (line 3) is missing.
As a result, an attacker could claim rewards even after the NFT has been unstaked.


\noindent\textbf{(6) Unauthorized Staking Asset Access (UAA):} 
This defect refers to the unauthorized transfer of staked assets or rewards belonging to other users.
For instance, this can be accomplished by directly modifying users’ balances, manipulating accounts to perform staking or unstaking operations, or unauthorizedly transferring staked tokens or rewards from any account.
An attacker could exploit these defects to access and steal users’ assets without their consent.

\vspace{-0.4cm}
\begin{figure}[h]
\setlength{\abovecaptionskip}{0.05cm}
\begin{lstlisting}[language=Solidity,mathescape]
function forceTransfer(address from, address to, uint _amount) public {
		userStakeAmount[from] = userStakeAmount[from].sub(_amount);
		userStakeAmount[to] = userStakeAmount[to].add(_amount);
    _standardTransfer(from, to, _amount); }
\end{lstlisting}
\caption{An example of Unauthorized Staking Asset Access defect.}
\label{fig:defect_uaa}
\end{figure}
\vspace{-0.3cm}

\textbf{Example:}
Figure~\ref{fig:defect_uaa} illustrates a simplified example of the \textit{Unauthorized Staking Asset Access} defect. 
In this example, the \textit{forgeTransfer} function facilitates the transfer of staking tokens from the $from$ account to a specified $to$ account (lines 2-4). However, the function lacks appropriate permission checks to verify whether the caller is the $from$ account or has the necessary authorization to access its tokens. 
As a result, an attacker can steal staking tokens from any account.

\section{Methodology}~\label{sec:method}
To aid smart contract developers in identifying logical defects within DeFi staking, we have developed a detection tool called SSR (\textbf{S}afeguarding \textbf{S}taking \textbf{R}ewards).

\subsection{Design Decision}
SSR utilizes static analysis to detect logical defects in DeFi staking contracts before their deployment on the blockchain, eliminating the need for transactional data or runtime information.
However, the inherent complexity of DeFi staking mechanisms introduces several technical challenges. 

\textbf{The Complexity of DeFi Staking Logic. }
A thorough understanding of the staking logic requires substantial expertise.
This complexity arises from several core functionalities, such as staking, claiming rewards, and unstaking, which distinguish DeFi staking contracts from other contract types.
Moreover, the staking logic can vary significantly across different contracts.
To address this challenge, SSR utilizes the code comprehension capabilities of LLMs to extract fundamental information related to DeFi staking.
The extracted information includes variables such as the staking amount, duration, and functionalities such as staking and claiming rewards.

\textbf{Identification of Transfer Amount Calculation. }
The calculation of transferred token amount, such as the amount of staking rewards, is a critical factor in DeFi staking contracts.
Accurately identifying its calculation logic presents two main challenges: first, locating the variables related to the transfer amount; second, deriving the corresponding calculation formula.
To address these challenges, SSR first identifies the variables that determine the transfer amount from LLM-extracted transfer statements. 
It then constructs a calculation dependency graph (CDG) based on these variables to identify the state variables involved in the transfer amount calculations.

\subsection{Overview}
\vspace{-0.3cm}
\begin{figure}[h]
    \centering
    \setlength{\abovecaptionskip}{0.05cm}
    \includegraphics[width=\linewidth]{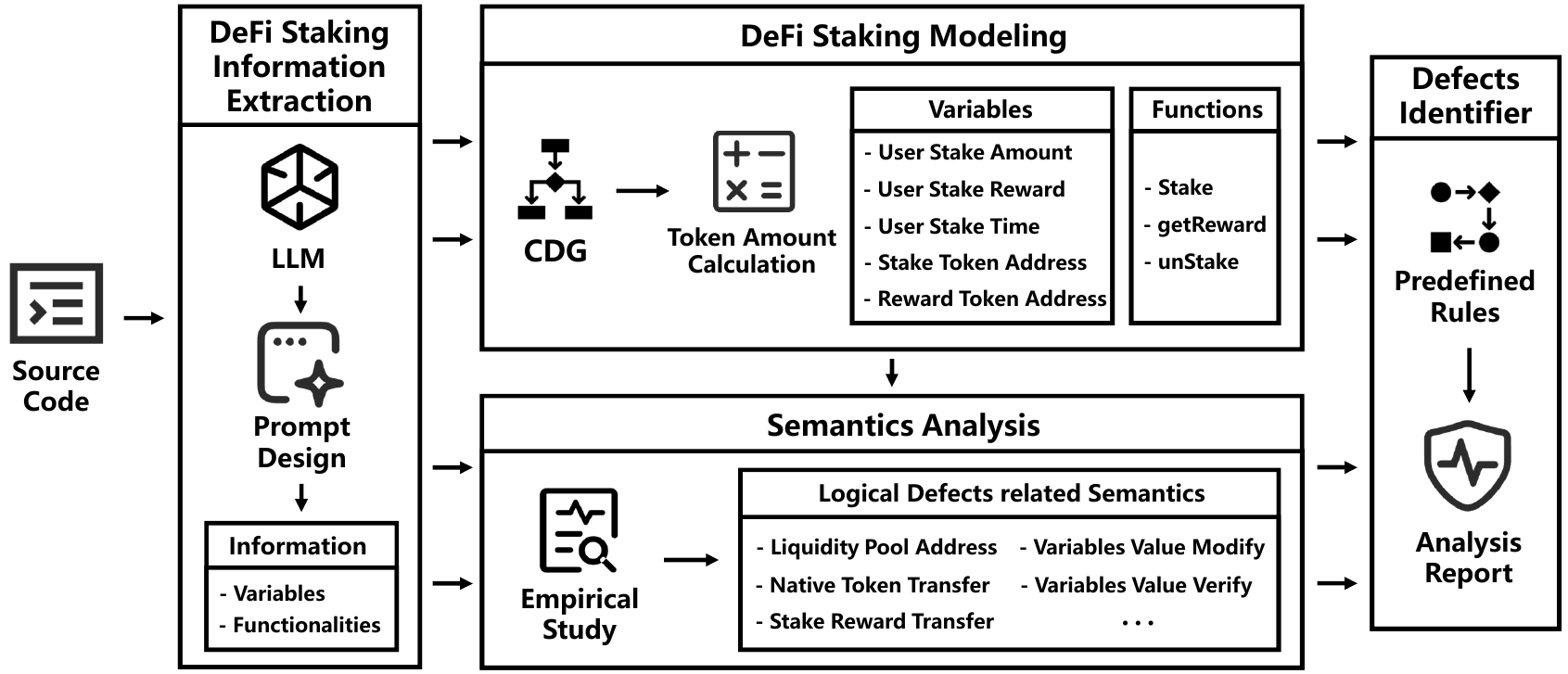}
    \caption{Overview of SSR}
    \label{fig:method_overview}
\end{figure}
\vspace{-0.3cm}

Figure~\ref{fig:method_overview} illustrates the overall framework of SSR. Given the source code of a DeFi staking contract, SSR first employs LLMs to extract fundamental information related to the DeFi staking logic (Section~\ref{subsec: method_infos}). 
Based on the extracted information, SSR constructs a DeFi staking model, encompassing relevant variables, functions, and the transfer amount calculation (Section~\ref{subsec: method_model}). 
Additionally, SSR extracts the semantics related to logical defects in DeFi staking based on the findings of prior empirical studies (Section~\ref{subsec: method_semantics}).
Ultimately, SSR detects six types of logical defects in DeFi staking based on the predefined rules (Section~\ref{subsec: method_identifier}).

\subsection{DeFi Staking Information Extraction}~\label{subsec: method_infos}
In this module, SSR leverages the open-source LLM DeepSeek-V3~\cite{deepseekai2024deepseekv3technicalreport} to extract fundamental information of DeFi staking, including relevant variables and functionalities.

Given the critical role of prompt design in determining the effectiveness of LLMs, we developed specialized prompt sets tailored to extract variables and functionalities relevant to DeFi staking. 
We adopted the prompt structure recommended by DeepSeek-V3~\cite{deepseek_prompt}, which comprises two components: the system prompt (SP) and the user prompt (UP). 
The following presents a detailed discussion of the prompt design process.

\textbf{System Prompt (SP).} 
The system prompt (SP) defines the role of the LLM as a smart contract auditor. 
To improve the stability and accuracy of the model’s responses, the LLM is instructed to generate multiple responses in the background and select the most frequent one.
Additionally, the output format is restricted to \textit{JSON}, excluding any explanatory text, thereby structuring the LLM’s output to facilitate subsequent modeling automation.
The SP used to extract variables and functionalities related to DeFi staking remains unchanged.

\begin{figure}[h]
    \centering
    \setlength{\abovecaptionskip}{0.05cm}
    \includegraphics[width=\linewidth]{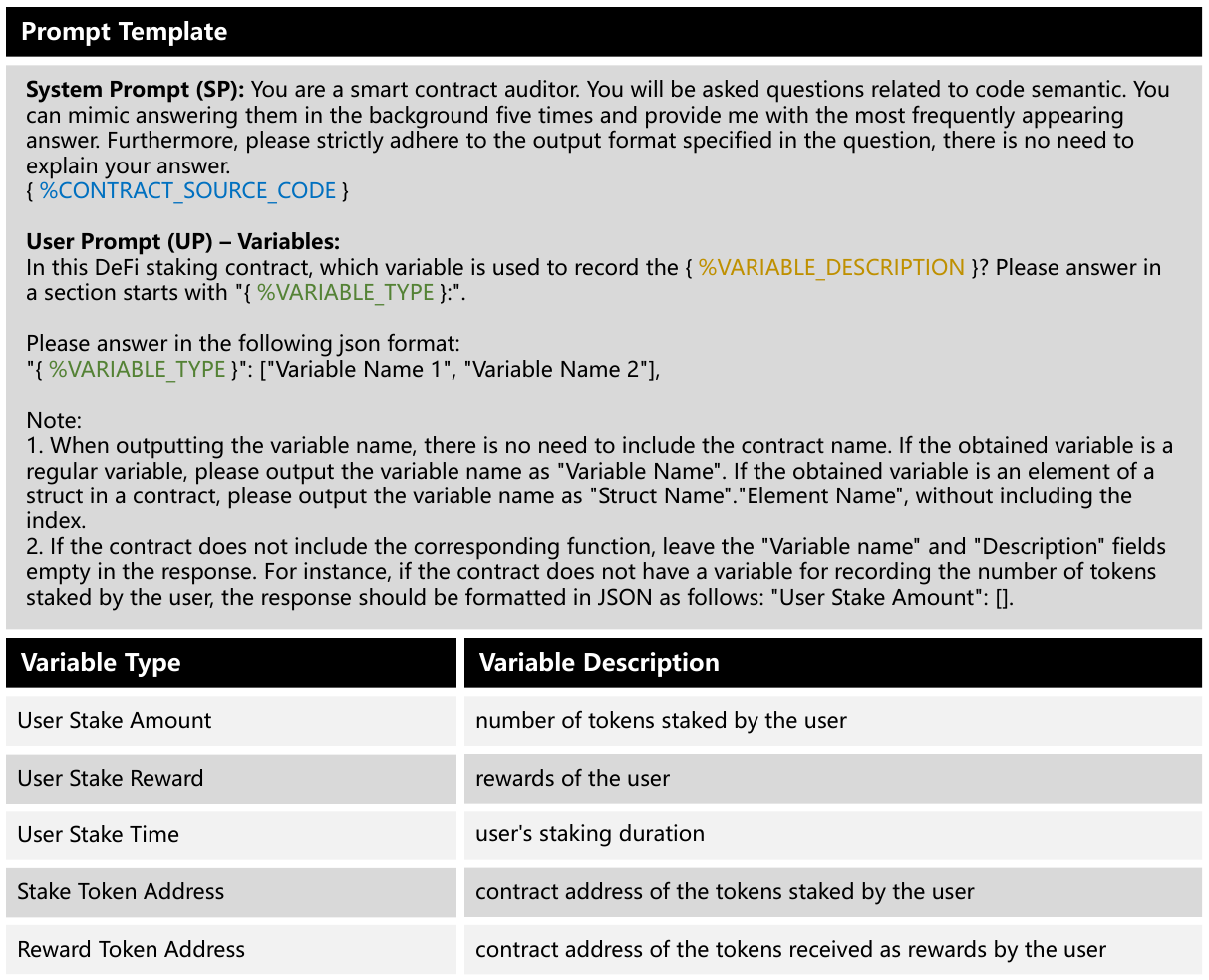}
    \caption{Prompt Template of DeFi Staking-related Variables}
    \label{fig:method_infos_prompt_var}
\end{figure}

\begin{figure}[h]
    \centering
    \setlength{\abovecaptionskip}{0.05cm}
    \includegraphics[width=\linewidth]{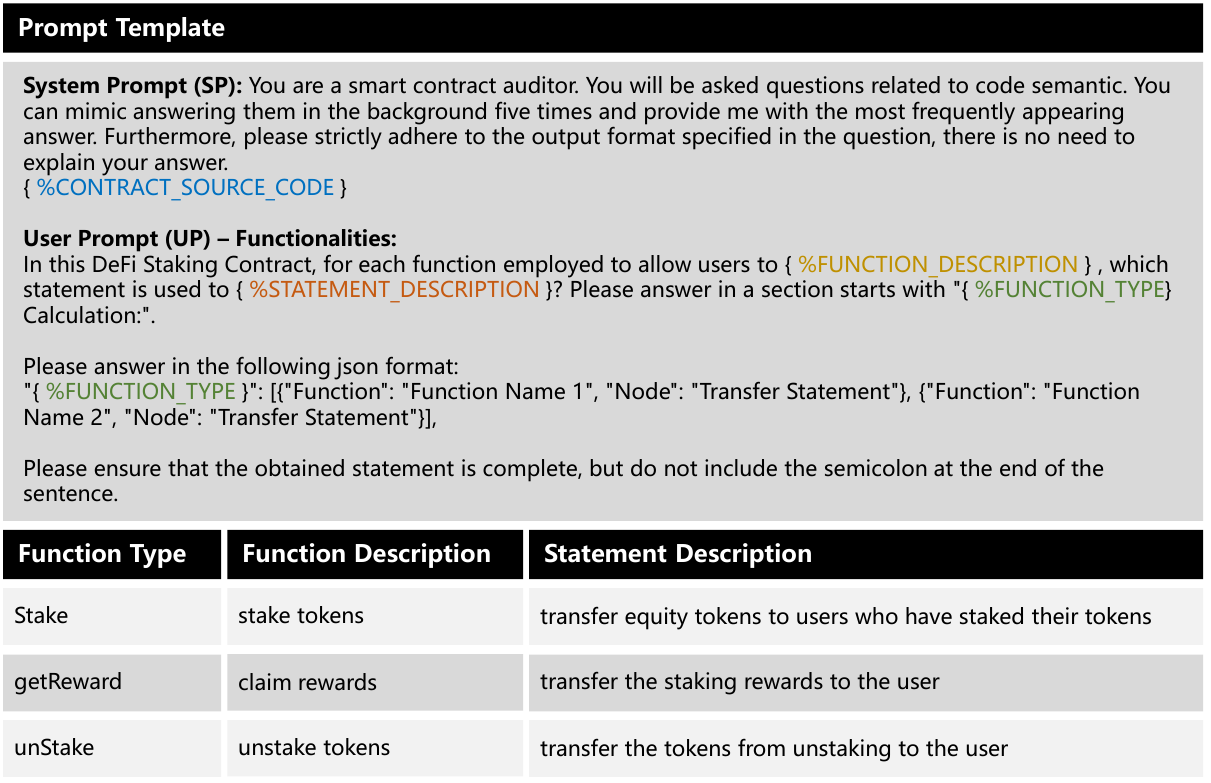}
    \caption{Prompt Template of DeFi Staking-related Functionalities}
    \label{fig:method_infos_prompt_func}
\end{figure}

\textbf{User Prompt (UP).} 
The UP is responsible for guiding the LLM in analyzing and extracting variables and functionalities related to DeFi staking.
The corresponding prompt templates are shown in Figures~\ref{fig:method_infos_prompt_var} and~\ref{fig:method_infos_prompt_func}, respectively. 

\vspace{-0.3cm}
\vspace{-0.3cm}

For variables, SSR identifies five types closely related to DeFi staking logic: \textit{User Stake Amount}, \textit{User Stake Reward}, \textit{User Stake Time}, \textit{Stake Token Address}, and \textit{Reward Token Address}. 
SSR first provides a description of each variable’s functionality and instructs the LLM to extract the corresponding variable names from the contract. The types and descriptions of these variables are illustrated in Figure~\ref{fig:method_infos_prompt_var}.
To reduce costs, SSR integrates the type names and descriptions of the five distinct variable types into a single prompt, enabling the simultaneous identification of these variables. 
Additionally, supplementary notes are appended to the end of the UP to further constrain the output format of the LLM, thereby enhancing the automation and accuracy of modeling.

For functionalities, SSR identifies three core operations in the DeFi staking: \textit{Stake}, \textit{getReward}, and \textit{unStake}. 
For each functionality, SSR first provides a description and guides the LLM in extracting the functions and statements that implement the corresponding logic.
The types and descriptions of these functionalities are illustrated in Figure~\ref{fig:method_infos_prompt_func}. 
For example, for the \textit{getReward} functionality, the LLM extracts the function that allows users to claim staking rewards, along with the specific statement that transfers the reward tokens to the user.
Similar to the approach used for variables, different functionality types are integrated into a single prompt to reduce costs, with the output format explicitly constrained at the end.

\vspace{-0.3cm}
\vspace{-0.3cm}

\subsection{DeFi Staking Modeling}~\label{subsec: method_model}
In this module, SSR constructs a DeFi staking model leveraging foundational information extracted by the LLM.
This model provides a critical basis for subsequent semantic analysis and defect detection.
The following details the model’s structure and the methodologies used in its construction.

\textbf{DeFi Staking Model Structure.}
The structure of the DeFi staking model consists of three components: DeFi staking-related variables, functions, and calculations of the transferred token amounts. 
Specifically, the variables include the \textit{User Stake Amount}, \textit{User Stake Reward}, \textit{User Stake Duration}, \textit{Stake Token Address}, and \textit{Reward Token Address}. 
The functions encompass three core operations in DeFi staking: staking, claiming rewards, and unstaking. 
The calculations of the transferred token amounts involve determining the number of equity tokens acquired through staking, the stake rewards earned, and the number of tokens received upon unstaking.

\textbf{The Model Construction Method.} 
SSR employs the smart contract analysis tool Slither~\cite{feist2019slither} to model the DeFi staking contract based on its source code.

In variable analysis, SSR first identifies relevant variables based on their names, which are extracted in the preceding module.
It then reconstructs the internal model of a two-layer data structure, which is commonly employed to store user stake information in DeFi contracts. 
A typical example is $userInfo.stakeAmount$, where $userInfo$ is a mapping from user addresses to a struct that contains individual staking data. Different keys within $userInfo$ convey distinct logical meanings, for instance, $userInfo.stakeTime$ denotes the timestamp when the user initiated staking. To accurately identify the variable that stores a user’s staked amount, it is crucial to determine both the mapping variable, $userInfo$, and the specific key, $stakeAmount$.

In function analysis, SSR first identifies functions related to the three core functionalities by their names, which are extracted in the preceding LLM-based module.
It then further refines and filters additional relevant functions by analyzing invocation relationships and evaluating whether they can be externally invoked by users.

In analyzing the calculation of the transferred token amount, SSR first locates the transfer statement and then identifies the variable that represents the transferred amount within it.
Subsequently, to identify the variables used for the transferred amount calculation, SSR constructs a calculation dependency graph (CDG).
The nodes in the CDG represent the variables that the transferred amount depends on, while the edges capture the dependencies among them.

\vspace{-0.3cm}
\begin{algorithm}
\fontsize{9}{12}\selectfont 
\caption{Construction of CDG}
\label{alg:cdg}
\begin{algorithmic}[1]
\REQUIRE Transferred amount variable $Var_{amount}$
\REQUIRE Smart Contract $contract$
\ENSURE Calculation Dependency Graph $CDG$

\STATE $CDG \leftarrow \emptyset$ 
\STATE $queue \leftarrow [Var_{amount}]$ 
\STATE $visited \leftarrow visited \cup \{Var_{amount}\}$

\WHILE{$queue$ is not empty}
    \STATE $Var_{cur} \leftarrow queue.dequeue()$
    \STATE $VarSet_{cal} \leftarrow \text{GetCalculationDepend}(Var_{cur}, contract)$
    \FOR{each $Var_{cal}$ in $Dep_{cal}$}
        \IF{$Var_{cal} \notin visited$}
            \STATE $queue.enqueue(Var_{cal})$
            \STATE $visited \leftarrow visited \cup \{Var_{cal}\}$
        \ENDIF
        \STATE $CDG.addCalcualtionEdge(Var_{cur}, Var_{cal})$
    \ENDFOR

    \STATE $VarSet_{con} \leftarrow \text{GetControlDepend}(Var_{cur}, contract)$
    \FOR{each $Var_{con}$ in $Dep_{con}$}
        \IF{$Var_{con} \notin visited$}
            \STATE $queue.enqueue(Var_{con})$
            \STATE $visited \leftarrow visited \cup \{Var_{con}\}$
        \ENDIF
        \STATE $CDG.addControlEdge(Var_{cur}, Var_{con})$
    \ENDFOR
    
\ENDWHILE

\end{algorithmic}
\end{algorithm}
\vspace{-0.3cm}


The algorithm for constructing the CDG is presented in Algorithm~\ref{alg:cdg}. Starting with a variable of the transferred token amount, the algorithm extracts all other variables on which it depends. This dependency analysis is recursively applied to the newly identified variables until no additional dependencies are found.
There are two types of edges \textit{$ E_{\text{cal}} $} and \textit{$ E_{\text{con}} $}, which represent the dependency between variable values in calculations and the control flow relationship of variable modification operations, respectively. 
These edges are determined by analyzing the variables read by the nodes in the DFG and CFG associated with the given variable.
Once the CDG is constructed, the state variables on which the transferred token amount depends are identified. 
Variables that do not contribute to detecting defects, such as local variables, are excluded.


\vspace{-0.4cm}
\begin{figure}[h]
\setlength{\abovecaptionskip}{0.05cm}
\begin{lstlisting}[language=Solidity,mathescape]
function pendingToken(address account) public view returns(uint) {
    unit _rewardPerToken = getPerTokenReward()
    unit rewardAmount = userLPStakeAmount[account]
        * (_rewardPerToken - userRewardPerTokenPaid[account]) / (1e18) + (userRewards[account]);
    return rewardAmount   }
function getReward() public {
    uint _reward = pendingToken(_msgSender());
    require(_reward > min, "The stake reward is less than the minimum obtainable reward.");
    userRewards[_msgSender()] = 0;
    if (_reward > 0) {
        _standardTransfer(address(this), _msgSender(), _reward);
        return ;  }  }
\end{lstlisting}
\caption{A concrete example for CDG Construction.}
\label{fig:method_cdg_exam}
\end{figure}
\vspace{-0.3cm}

Taking the contract in Figure~\ref{fig:method_cdg_exam} as a concrete example, when a user calls function $getReward()$ (line 6-10) to claim staking rewards, the contract transfers an amount, $\_reward$, to the user (line 11). 
Since $\_reward$ is calculated using the function $pendingToken()$ (lines 1-5), depending on $\_rewardPerToken$, $userLPStakeAmount$, and others, the \textit{$ E_{\text{cal}} $} between $\_reward$ and these variables is extracted. 
Furthermore, since the transfer is conditioned on the $\_reward$ exceeding the minimum threshold $min$ (line 8), the \textit{$ E_{\text{con}} $} between $\_reward$ and $min$ is extracted.
The process then recursively expands to capture additional dependencies, such as $\_rewardPerToken$, which is calculated in the $getPerTokenReward()$ function (line 2).

Additionally, since the balance of native tokens held by the contract is stored in a special variable, $this.balance$, SSR performs an additional analysis to determine whether the transfer amount depends on this token balance.

Finally, by constructing the DeFi Staking model described above, SSR extracts information related to the DeFi Staking logic. The names and descriptions of the extracted logic are presented in Table~\ref{tab:method_model_infos}.

\vspace{-0.4cm}
\begin{table}[h] 
    \centering
    \setlength{\abovecaptionskip}{0.05cm}
        \caption{Extracted DeFi Staking Logic}
        \resizebox{\linewidth}{!}{
        \begin{tabular}{p{2.7cm}|p{5.8cm}}
            \hline
            \textbf{Name} & \textbf{Description} \\
            \hline
            \textit{Reward($var$)} & Variable $ var $ denotes the DeFi staking reward. \\
            \hline
            \textit{StakeTime($var$)} & Variable $ var$ denotes the duration of time during which the user holds a stake in the asset. \\
            \hline
            \textit{Amount($var$)} & Variable $ var$ denotes the number of tokens that the user has staked. \\
            \hline
            \textit{unStake($func$)} & Function $ func $ is used to transfer the tokens from unstaking to the user. \\
            \hline
            \textit{CalDepend($var_{1}$, $var_{2}$)} & The value of the variable $ var_{1} $ is calculated based on the value of the variable $ var_{2} $. \\
            \hline
            \textit{DependonBalance($var$)} & The value of the variable $var$ is calculated based on the balance of native tokens in the contract.  \\
            \hline

        \end{tabular}}
    \label{tab:method_model_infos}
\end{table}
\vspace{-0.3cm}

\subsection{Semantics Analysis}~\label{subsec: method_semantics}
In this module, SSR extracts semantics related to logical defects in DeFi staking based on the DeFi staking model. This includes semantics concerning the meanings of variables and the behaviors of functions. The names and descriptions of extracted semantics are presented in Table~\ref{tab:method_semantics}.

\vspace{-0.4cm}
\begin{table}[h] 
    \centering
    \setlength{\abovecaptionskip}{0.05cm}
        \caption{Semantics related to logical defects in DeFi staking}
        \resizebox{\linewidth}{!}{
        \begin{tabular}{p{2.3cm}|p{6.3cm}}
            \hline
            \textbf{Name} & \textbf{Description} \\
            \hline
            \textit{lpPool($con$)} & Variable $ con $ denotes an external liquidity pool contract. \\
            \hline
            \textit{ModifyVar($func$, $var$)} & The value of the variable $ var $ is modified within the function $ func $. \\
            \hline
            \textit{VerifyVar($func$, $var$)} & The execution of certain logic within the function $ func $ requires the verification of variable $ var $. \\
            \hline
            \textit{NaTokenTrans($func$, $to$)} & In function $ func $, the native token of the contract is transferred to the address $ to $. \\
            \hline
            \textit{RewardTrans($func$, $from$, $to$)} & In function $ func $, the staking reward of the address $ from $ is transferred to the address $ to $. \\
            \hline
            \textit{permissCheck($func$, $addr$)} & The execution of certain logic within the function requires verifying whether the function caller is the address $addr$ or possesses authorization for $addr$.  \\
            \hline

        \end{tabular}}
    \label{tab:method_semantics}
\end{table}
\vspace{-0.3cm}

SSR analyzes the semantics of liquidity pool contracts by examining their features, including the presence of functions such as mint and burn, as well as the ability to determine the ratio between two distinct tokens within the pool. These features are inferred through the analysis of widely adopted third-party liquidity pool contracts, including Uniswap~\cite{uniswap} and PancakeSwap~\cite{pancakeswap}.


$ModifyVar$ and $VerifyVar$ indicate whether a function modifies or verifies the value of a specific variable, respectively. 
For $VerifyVar$, SSR determines the verification of a variable’s value by examining specific nodes in the CFG, such as modifiers and require statements. 
The logic controlled by these nodes depends on verifying the value of the variable.
For $ModifyVar$, SSR determines the modification of a variable’s value by checking whether any function performs a write operation on the variable.
By modeling the two-layer data structure of variables in the \textit{DeFi Staking Modeling} module, SSR can identify functions that modify or verify these variables, as well as the special keys that are modified or verified.

$NaTokenTrans$ and $RewardTrans$ represent the transfer of native tokens and staking rewards within the functions, respectively. 
SSR identifies token transfer based on the naming and parameter conventions defined in the ERC standards~\cite{erc20, erc721} and determines whether the transferred token is a native token or a reward token. 
Specifically, it recognizes native tokens by detecting the special variable $this.balance$ and identifies staking rewards using the DeFi staking model.

$permissCheck$ represents the permission check performed within a function, including determining whether the caller is a specific address and verifying whether the caller has the necessary permissions, such as authorization to transfer another user’s tokens.
SSR first extracts the control-flow nodes that govern a function’s execution from the contract’s CFG, including special nodes like modifiers and require statements. 
It then analyzes whether these nodes contain logic that verifies if the function invoker corresponds to a specific address. 
If such logic is found, SSR considers it a permission check within the function.

\subsection{Defects Identifier}~\label{subsec: method_identifier}
In this module, SSR leverages the DeFi staking model and semantic features to identify six distinct types of logical defects in DeFi staking. The following provides detailed explanations of the detection methods for each defect type.

\noindent\textbf{(1) Staking Logical Variables Manipulation (SVM)}
SSR identifies two scenarios that could lead to this defect: dependence on ordinary variables and the native token balance.
The detection logic for both scenarios is outlined in Rule~\ref{equa:method_identifer_svm1} and Rule~\ref{equa:method_identifer_svm2}, respectively.
Rule~\ref{equa:method_identifer_svm1} identifies a defect when staking rewards depend on a variable that arbitrary external accounts can modify.
Rule~\ref{equa:method_identifer_svm2} identifies a defect when staking rewards depend on the native token balance, which arbitrary external accounts can transfer.
In addition to the scenario involving staking rewards, SSR applies similar logic to detect cases related to tokens deposited during staking and unstaking.
\begin{equation}
\resizebox{0.9\hsize}{!}{
$
isSVM = \begin{aligned}
&Reward(re) \land CalDepend(re, var)  \land \\ & ModifyVar(func, var)  \land \neg permissCheck(func, \_)
\end{aligned}
$
}
\label{equa:method_identifer_svm1}
\end{equation}
\begin{equation}
\resizebox{0.9\hsize}{!}{
$
isSVM = \begin{aligned}
&Reward(re) \land DependonBalance(re)  \land \\ & NaTokenTrans(func, \_)  \land \neg permissCheck(func, \_)
\end{aligned}
$
          }
\label{equa:method_identifer_svm2}
\end{equation}

\noindent\textbf{(2) Rewards without Timedelay (RT)}
The logic for detecting this defect is defined in Rule~\ref{equa:method_identifer_rt}. 
A defect is identified when the calculation of staking rewards does not rely on any variable associated with the staking duration ($\neg StakeTime(var)$). 
In determining whether a variable is related to staking duration, SSR considers not only relevant variables in the DeFi staking model but also special time-related variables in Solidity~\cite{solidity} smart contracts, such as $block.timestamp$.
\begin{equation}
\resizebox{0.9\hsize}{!}{
$isRT = Reward(re) \land \neg StakeTime(var) \land CalDepend(re, var)$
          }
\label{equa:method_identifer_rt}
\end{equation}

\noindent\textbf{(3) Single Liquidity Pool Reliance (SLR)}
The logic for detecting this defect is defined in Rule~\ref{equa:method_identifer_slr}. A defect is identified when the staking rewards depend solely on a single liquidity pool. 
In Rule~\ref{equa:method_identifer_slr}, $ \left\{ \mathit{con} \mid CalDepend(re, \mathit{con}), lpPool(\mathit{con}) \right\} $ represents the set of liquidity pool contracts on which the calculation of staking reward depends.
\begin{equation}
\resizebox{0.9\hsize}{!}{
$
isSLR = \begin{aligned}
&Reward(re) \land \# Pool == 1  \land \\ &Pool = \left\{ con \mid CalDepend(re, con),\
      lpPool(con) \right\}
\end{aligned}
$
          }
\label{equa:method_identifer_slr}
\end{equation}

\noindent\textbf{(4) Omission in Status Update (OSU)}
This defect involves three scenarios: (i) whether the staked amount and duration are updated in the stake function; (ii) whether the staking rewards and duration are updated in the claim reward function; and (iii) whether the staked amount is updated in the unstake function.
Taking the status update in the \textit{unStake} function as an example, the detection logic is shown in Rule~\ref{equa:method_identifer_osu}. If the staked tokens amount ($Amount(a)$) is not updated in the \textit{unStake} function ($unStake(func)$), a defect is considered to exist.
\begin{equation}
\resizebox{0.95\hsize}{!}{
$isOSU = unStake(func) \land Amount(a) \land ModifyVar(func, a)$
          }
\label{equa:method_identifer_osu}
\end{equation}

\noindent\textbf{(5) Unsafe Verification (UV)}
This defect involves two scenarios: whether the staked amount is verified in the unstake function, and whether the staking reward is verified in the claim reward function. Taking the unstake function as an example, the detection logic is shown as Rule~\ref{equa:method_identifer_uv}. If the amount of staked tokens is not verified ($\neg verifyVar(func, a)$) before the token transfer is executed, a defect is considered to exist.
\begin{equation}
\resizebox{0.9\hsize}{!}{
$isUV = unStake(func) \land Amount(a) \land \neg verifyVar(func, a)$
          }
\label{equa:method_identifer_uv}
\end{equation}

\noindent\textbf{(6) Unauthorized Staking Asset Access (UAA)}
This defect involves two scenarios: transferring the user’s staked token or rewards through token transfer, or directly modifying the token balance. 
Taking staking rewards as an example, the detection logic for both scenarios is in Rule~\ref{equa:method_identifer_uaa1} and Rule~\ref{equa:method_identifer_uaa2}, respectively.
Rule~\ref{equa:method_identifer_uaa1} states that if a function permits the transfer of rewards from the address $from$ to another address, but fails to validate whether the caller is $from$ or check for the appropriate permissions, a defect is considered to exist.
Rule~\ref{equa:method_identifer_uaa2} states that if a function can directly modify the staking rewards without permission checks, a defect is considered to exist.
\begin{equation}
\resizebox{0.9\hsize}{!}{
$isUAA = RewardTrans(func, from, $ \_$) \land \neg perCheck(func, from) $
          }
\label{equa:method_identifer_uaa1}
\end{equation}
\begin{equation}
\resizebox{\hsize}{!}{
$isUAA = Reward(re) \land modifyVar(func, re) \land \neg perCheck(func, from)$
          }
\label{equa:method_identifer_uaa2}
\end{equation}

\section{Evaluation}
\subsection{Evaluation Setup}
\textbf{Research Questions.} 
We focus on the following three research questions.
\begin{itemize}
    \item \textbf{RQ1:} How effective is SSR in detecting logical defects in DeFi staking contracts?
    \item \textbf{RQ2:} What is the prevalence of logical defects in real-world DeFi staking contracts?
    \item \textbf{RQ3:} How accurate is SSR in modeling the logic associated with DeFi staking in contracts?
\end{itemize}

\textbf{Dataset.}
To address the research questions outlined above, we constructed two datasets. 
First, to assess the effectiveness of SSR in detecting logical defects in DeFi staking, we created a ground truth dataset consisting of 40 DeFi staking contracts, collected based on the security incidents and audit reports. 
Second, to investigate the prevalence of logical defects in real-world DeFi staking contracts, we compiled a large-scale dataset comprising 15,992 such contracts. 
The details of these two datasets are described below.

For the ground truth dataset, we first collected open-source smart contracts associated with the security incidents and audit reports analyzed in the previous empirical study. 
We then manually verified whether each contract qualified as a DeFi staking contract and whether it could be successfully compiled. As a result, we obtained 40 DeFi staking contracts, comprising 21 positive cases and 19 negative cases.

\vspace{-0.4cm}
\begin{table}[h] 
    \centering
    \setlength{\abovecaptionskip}{0.05cm}
        \caption{The dataset scale at each preprocessing step}
        \resizebox{\linewidth}{!}{
        \begin{tabular}{|l|c|c|c|c|}
            \hline
                           & \textbf{Original}              & \textbf{Keyword Filtered}    & \textbf{Deduplicated}        & \textbf{LLM Filtered}       \\ \hline
            \textbf{Scale} & \multicolumn{1}{r|}{1,030,719} & \multicolumn{1}{r|}{137,642} & \multicolumn{1}{r|}{69,444} & \multicolumn{1}{r|}{15,992} \\ \hline
        \end{tabular}
        }
    \label{tab:eva_dataset}
\end{table}
\vspace{-0.3cm}

For the large-scale dataset, we initially collected 1,030,719 open-source smart contracts from nine blockchains~\cite{etherscan, bscscan, celo, tron, polygon, arbitrum, avalanche, fantom, optimism}, sourced from a widely used repository~\cite{smart_contract_sanctuary}. 
Data preprocessing was carried out in three stages. 
First, an initial filter was applied using the keyword ``stak'' to identify potential DeFi staking contracts. 
Next, duplicates were removed, and those that were not compilable were discarded. 
In the final stage, the LLM DeepSeek-V3 was used to verify whether these contracts are indeed DeFi staking contracts.
As a result, we obtained 15,992 real-world DeFi staking contracts. The dataset size at each preprocessing step is presented in Table~\ref{tab:eva_dataset}.

The experiment was executed on a server running Ubuntu 22.04.1 LTS equipped with 32 Intel(R) Xeon(R) Platinum 8360H CPUs (3.00GHz) and 64 GB of memory.

\subsection{RQ1: Effectiveness of SSR in the Ground Truth Dataset} ~\label{subsec: eva_rq1}
To answer RQ1, we applied SSR to the ground truth dataset, which consists of 40 DeFi staking contracts. 
The results are summarized in Table~\ref{tab:eva_groundTruth}, which reports the number of each type of logical defect (Incs), along with the counts of true positives (TP), false positives (FP), and false negatives (FN).
Precision, recall, and F1-score for identifying each defect were calculated using the following formulas: $\frac{\#TP}{\#TP + \#FP} \times 100\%$, $\frac{\#TP}{\#TP + \#FN} \times 100\%$, and $\frac{2\times Recall\times Precision}{Recall+Precision}$. 
Additionally, we calculated the overall evaluation metrics to assess the effectiveness of SSR. The overall precision is computed as $ \frac{{\textstyle \sum_{i=1}^{n}p_{c_i} \times \left | c_i \right | }}{ {\textstyle \sum_{i=1}^{n} \left | c_i \right | } } $, where $ p_{c_i} $ denotes the precision for detecting defect type $ i $, and $ |c_i| $ is the number of contracts with defect $ i $. 
The overall recall and F1-score were computed using analogous formulas by substituting the respective metric for $ p_{c_i} $. 
This evaluation approach is consistent with methodologies employed in prior studies~\cite{lin2024crpwarner, yang2024hyperion}.

\vspace{-0.4cm}
\begin{table}[h] 
    \centering
    \setlength{\abovecaptionskip}{0.05cm}
        \caption{Detection Result in ground truth dataset}
        \resizebox{\linewidth}{!}{
        \begin{tabular}{|l|l|l|l|l|l|l|l|}
\hline
\textbf{Type}                        & \multicolumn{1}{c|}{\textbf{\# Incs}} & \textbf{\# TP} & \textbf{\# FP} & \textbf{\# FN} & \multicolumn{1}{c|}{\textbf{Prec (\%)}} & \multicolumn{1}{c|}{\textbf{Rec (\%)}} & \multicolumn{1}{c|}{\textbf{F1 (\%)}} \\ \hline
SVM                                  & 1                                  & 1              & 0              & 0              & 100.00                                     & 100.00                                    & 100.00                                   \\ \hline
RT                                   & 5                                  & 4              & 0              & 1              & 100.00                                     & 80.00                                     & 88.89                                 \\ \hline
SLR                                  & 3                                  & 2              & 0              & 1              & 100.00                                     & 66.67                                  & 80.00                                    \\ \hline
OSU                                  & 5                                  & 4              & 0              & 1              & 100.00                                     & 80.00                                     & 88.89                                 \\ \hline
UV                                   & 6                                  & 6              & 2              & 0              & 75.00                                   & 100.00                                    & 85.71                                    \\ \hline
UAA                                  & 8                                  & 7              & 0              & 1              & 100.00                                     & 87.50                                  & 93.33                                 \\ \hline
\multicolumn{1}{|c|}{\textbf{Total}} & /                                  & /              & /              & /              & 92.31                                   & 87.92                                   & 88.85                                 \\ \hline
\end{tabular}
        }
    \label{tab:eva_groundTruth}
\end{table}
\vspace{-0.3cm}

The experimental results indicate that SSR achieves an overall precision of 92.31\%, a recall of 87.92\%, and an F1-score of 88.85\% on the ground truth dataset.

Since the causes of FPs and FNs are consistent across both the ground truth and large-scale datasets, their analysis is presented collectively in response to RQ2.

\subsection{RQ2: Prevalence of Logical Defects in DeFi Staking} ~\label{subsec: eva_rq2}
To answer RQ2, we applied SSR to a large-scale dataset consisting of 15,992 unlabeled DeFi staking contracts. The detection results are presented in Table~\ref{tab:eva_largeScale}. The second and third columns in the table display the number of logical defects detected by SSR for each type, along with the corresponding percentage. The results indicate that 3,557 (22.24\%) of the contracts in the dataset contain at least one type of defect.

To assess the effectiveness of SSR in detecting logical defects in DeFi staking in a large-scale dataset, we employed random sampling combined with manual verification. Specifically, to evaluate the precision of SSR, we conducted random sampling at a 95\% confidence level with a 10\% margin of error for each type of logical defect. This sampling methodology is consistent with prior studies~\cite{lin2024crpwarner, lin2024definition, zhang2024demystifying}. The sample sizes for each defect type are provided in the fourth column of Table~\ref{tab:eva_largeScale}.
Given the rarity of certain logical defects, we performed separate random samplings for each defect type, rather than sampling directly from the entire dataset. 
For instance, contracts identified by SSR as exhibiting the \textit{Single Liquidity Pool Reliance} (SLR) defect account for only 0.28\% of the dataset. A random sample drawn from the entire dataset may fail to capture these rare defects, which could hinder an evaluation of SSR’s performance for this defect.

\vspace{-0.4cm}
\begin{table}[h] 
    \centering
    \setlength{\abovecaptionskip}{0.05cm}
        \caption{Detection Result in large-scale dataset}
        \resizebox{\linewidth}{!}{
\begin{tabular}{|l|l|l|l|l|l|l|}
\hline
\multicolumn{1}{|c|}{\textbf{Type}}                        & \multicolumn{1}{c|}{\textbf{\# Defects}} & \textbf{Per (\%)} & \textbf{\# Sample} & \textbf{\# TP} & \multicolumn{1}{c|}{\textbf{\# FP}} & \multicolumn{1}{c|}{\textbf{Prec (\%)}} \\ \hline
SVM                                  &                           430               &        2.69           &          79          &          68      &               11                      &         86.08                                \\ \hline
RT                                   &         516                                 &       3.23            &           82         &        79        &             3                        &        96.34                                \\ \hline
SLR                                  &             44                             &        0.28           &      31              &        29        &          2                           &         93.55                                \\ \hline
OSU                                  &                  889                        &         5.56          &         87           &       79         &          8                           &           86.21                              \\ \hline
UV                                   &                   1,315                       &          8.22         &     90               &        80        &            10                         &            88.89                            \\ \hline
UAA                                  &            846                              &         5.29         &             87       &        79        &                 8                    &          90.81                               \\ \hline
\multicolumn{1}{|c|}{\textbf{Total}} &            3,557                             &        22.24           &           /         &         /       &                  /                   &           89.41                              \\ \hline
\end{tabular}
        }
    \label{tab:eva_largeScale}
\end{table}
\vspace{-0.3cm}

We manually verified the sampled datasets to determine the number of true positives (TP) and false positives (FP) detected by SSR for each type of logical defect, along with the corresponding precision (Prec), as presented in Table~\ref{tab:eva_largeScale}. The formulas used to calculate the precision for each defect type, as well as the overall precision, are consistent with those employed in the analysis of RQ1. The experimental results indicate that SSR achieved an overall precision of 89.41\% in detecting logical defects in the large-scale dataset.

\vspace{-0.4cm}
\begin{table}[h] 
    \centering
    \setlength{\abovecaptionskip}{0.05cm}
        \caption{False negatives of SSR in large-scale dataset}
        \resizebox{0.9\linewidth}{!}{
\begin{tabular}{|l|l|l|}
\hline
\multicolumn{1}{|c|}{\textbf{Defect Type}}                       & \multicolumn{1}{c|}{\textbf{\# FN}} & \textbf{FN rate (\%)} \\ \hline
Staking Logical Variables Manipulation (SVM)                                  &       1                              &            1.04           \\ \hline
Rewards without Timedelay (RT)                                   &          2                           &          2.08             \\ \hline
Single Liquidity Pool Reliance (SLR)                                  &       2                              &          2.08             \\ \hline
Omission in Status Update (OSU)                                  &       3                              &            3.13           \\ \hline
Unsafe Verification (UV)                                  &            2                         &          2.08             \\ \hline
Unauthorized Staking Asset Access (UAA)                                &       4                              &         4.17              \\ \hline
\multicolumn{1}{|c|}{\textbf{Total}} &                     11                &       11.46                \\ \hline
\end{tabular}
        }
    \label{tab:eva_largeScale_fn}
\end{table}
\vspace{-0.3cm}

In addition, to validate the presence of logical defects that were not identified by SSR, we performed random sampling at a 95\% confidence level with a 10\% confidence interval on contracts for which SSR produced negative detection results. We randomly sampled and manually verified 96 such contracts in which no logical defects were detected by SSR. The experimental results are presented in Table~\ref{tab:eva_largeScale_fn}. These results indicate that SSR exhibits a false negative rate of 11.46\% on the large-scale dataset.

\textbf{False Positives (FPs).} 
For RT, OSU, and UV, the primary causes of FPs arise from limitations in the model’s ability to capture complex staking logic during the DeFi staking modeling process.

\vspace{-0.4cm}
\begin{figure}[h]
\setlength{\abovecaptionskip}{0.05cm}
\begin{lstlisting}[language=Solidity,mathescape]
mapping (address => uint) public stakingTime; //used for the unstaking locktime
mapping (address => uint) public firstTime; //used for the APY boost
mapping (address => uint) public lastClaimedTime; //used for the claiming locktime
mapping (address => uint) public progressiveTime; //used for the claiming locktime
\end{lstlisting}
\caption{An example of Unauthorized Staking Asset Access defect.}
\label{fig:eva_errorExample}
\end{figure}
\vspace{-0.3cm}

For example, as illustrated in Figure~\ref{fig:eva_errorExample}, several variables related to users’ staking times are defined.
Although these variables share the same data structure, their semantics and functionalities differ. 
Specifically, the variable $stakeTime$ represents the time threshold at which a user can withdraw their staked assets; $firstTime$ logs the timestamp of a user’s initial staking activity, which is used to calculate bonuses for the annual percentage yield (APY); $lastClaimedTime$ records the most recent time staking rewards were claimed, which is used to calculate the staking reward; and $progressiveTime$ is used to compute progressive staking rewards.

SSR struggles to accurately identify the complex logic surrounding staking time in DeFi staking models.
As a result, it fails to determine the staking time variables on which staking rewards depend.
Furthermore, it cannot distinguish between which time variables need to be updated following a staking reward claim and which do not.
In this case, $lastClaimedTime$  needs to be updated after a staking reward claim, while $firstTime$ does not. However, SSR incorrectly assumes that $firstTime$ must also be updated, resulting in a false OSM defect report.

For SVM, SLR, and UAA, the primary source of FPs stems from inaccuracies in pattern-based permission analysis and variable identification. For example, due to incorrect recognition of permission controls, SSR mistakenly classifies functions intended for adjusting project strategies, such as modifying the transaction fee rate, which should be restricted to project developers, as callable by any external account. Because staking rewards depend on the transaction fee rate, this misidentification results in a false SVM defect report.

\textbf{False Negatives (FNs).} 
For RT, OSU, and UV, the primary cause of FNs lies in the limitations of the LLM, which lead to the incorrect identification of DeFi staking-related variables. For instance, if the LLM mistakenly identifies a variable unrelated to stake time as the stake time, it could result in the omission of an RT defect.

For SLR, SVM, and UAA, the primary cause of FNs stems from errors or omissions in the control flow graph (CFG) and data flow graph (DFG) generated by Slither. For example, inaccuracies in the DFG may hinder the identification of all variables involved in the staking reward calculation. This could result in missing a variable that may be subject to external manipulation, leading to the omission of an SVM defect.

Since the complex staking calculation is the primary cause of FPs and FNs, we performed an analysis of the prevalence of these complex calculation patterns in real-world contracts. A random sample was taken at a 95\% confidence level with a 10\% confidence interval from the large-scale dataset. The results show that 5.21\% (5/96) of the samples contain the complex patterns, which SSR could not identify.

\subsection{RQ3: Accuracy of the DeFi Staking Model}~\label{subsec: eva_rq3}
Since the accuracy of the DeFi staking model directly impacts the effectiveness of SSR in detecting logical defects, we further validated the accuracy of the DeFi staking model by manually modeling the contracts in the ground truth dataset and comparing them with the models constructed by SSR.

The accuracy of the DeFi staking model is calculated by the average of three component accuracies: variables, functions, and calculations of transferred token amounts. And the accuracy of each component is determined by comparing the modeled variables and functions with the ground truth lists obtained through manual analysis. 
The calculation formula is as follows: $M_{\text{total}} = \frac{M_{\text{Var}} + M_{\text{Func}} + M_{\text{Cal}}}{3} $, where $ M_{\text{Var}} $, $ M_{\text{Func}} $, and $ M_{\text{Cal}} $, represent the evaluation metrics of modeling accuracy for DeFi staking-related variables, functions, and token amount calculations, respectively. 
$ M_{\text{total}} $ denotes the overall modeling accuracy for the DeFi staking contract. 
These metrics include precision, recall, and F1-score.

Taking the accuracy of the model for variables as an example, five distinct types of variables are identified in the model. The accuracy of each variable type is evaluated individually, and their average is calculated to determine the overall accuracy of the variable modeling.
A similar approach is applied to assess the accuracy of the other components.

\vspace{-0.3cm}
\begin{table}[h] 
    \centering
    \setlength{\abovecaptionskip}{0.05cm}
        \caption{The accuracy of DeFi Staking Modeling}
        \resizebox{0.9\linewidth}{!}{
\begin{tabular}{|l|l|l|l|}
\hline
\textbf{}                            & \multicolumn{1}{c|}{\textbf{Precision (\%)}} & \textbf{Recall (\%)} & \textbf{F1-Score (\%)} \\ \hline
Variables                            & 71.75                                        & 72.08                & 71.17                  \\ \hline
Functions                            & 90.72                                        & 91.46                & 90.38                  \\ \hline
Calculations                         & 91.25                                        & 94.05                & 91.00                  \\ \hline
\multicolumn{1}{|c|}{\textbf{Total}} & 84.57                                        & 85.87                & 84.18                  \\ \hline
\end{tabular}
        }
    \label{tab:eva_model}
\end{table}
\vspace{-0.3cm}

The accuracy of the DeFi Staking model is validated using the ground truth dataset, and the results are summarized in Table~\ref{tab:eva_model}. 
The DeFi Staking Modeling module of SSR achieved an overall precision of 84.57\%, recall of 85.87\%, and F1-score of 84.18\%. 
These results indicate that, while SSR may encounter difficulties in accurately identifying certain complex DeFi staking logic, it remains effective in modeling the majority of staking logic with high accuracy.

\vspace{-0.3cm}
\begin{table}[h] 
    \centering
    \setlength{\abovecaptionskip}{0.05cm}
        \caption{The accuracy of DeFi Staking Model using different LLM}
        \resizebox{0.9\linewidth}{!}{
\begin{tabular}{|l|l|l|l|}
\hline
\textbf{Model} & \textbf{Precision (\%)} & \textbf{Recall (\%)} & \textbf{F1-Score (\%)} \\ \hline
DeepSeek-v3    & \textbf{84.57}          & \textbf{85.87}       & \textbf{84.18}         \\ \hline
DeepSeek-r1    & 80.52                   & 79.94                & 78.53                   \\ \hline
GPT-4o         & 77.86                   & 79.20                & 77.10                  \\ \hline
\end{tabular}
        }
    \label{tab:eva_llm_compare}
\end{table}
\vspace{-0.3cm}

To evaluate the impact of various LLMs on DeFi staking modeling, Table~\ref{tab:eva_llm_compare} presents the accuracy of DeFi staking models constructed using several commonly used LLMs. 
The results show no significant differences among the different LLMs.
The primary factor influencing accuracy variations among LLMs is their differing ability to extract variables related to DeFi staking. In contrast, the accuracy of modeling both DeFi staking functions and the calculation of transferred amounts shows minimal variation.
Therefore, considering both recognition accuracy and usage costs, SSR selects DeepSeek-V3 for DeFi staking modeling.

\section{Threats to Validity}
\subsection{External Validity.}
The \textit{Defects Identifier} module of SSR is designed to detect six predefined types of logical defects. 
However, a limitation remains when new defect types are not included in the existing categories.
To address this issue, we collected a broad range of real-world security incidents and audit reports related to DeFi staking. 
These materials were analyzed and categorized using an open card sorting method to improve the comprehensiveness of the classification. 
Moreover, the SSR framework is inherently extensible, enabling future research to incorporate additional defect types.

In the evaluation phase, random sampling was used to assess the effectiveness of SSR, which could introduce sample bias. To mitigate this concern, we adopted a confidence-based random sampling method to reduce potential bias. This strategy has been employed in prior studies~\cite{yang2023definition, zhang2024demystifying} and offers a more statistically grounded approach to evaluation.

\subsection{Internal Validity.}
When constructing the ground truth dataset and labeling the data obtained through random sampling, we employed a manual labeling approach, which could be prone to errors. 
To mitigate this risk, the labeling was performed by two researchers, followed by a double-checking process. 
Both researchers were involved in the empirical study and possess significant expertise in logical defects in DeFi staking.

\section{Related Work}
\subsection{DeFi Staking Analysis}
Existing studies on DeFi staking primarily focus on the tokenomics, examining the trading and distribution of staked tokens. 
Based on these analyses, researchers can predict the future distribution of tokens and develop strategies.

Lin et al.~\cite{cong2025tokenomics} developed a continuous-time economic model featuring a tokenized digital network and analyzed how DeFi staking platforms manage their lifecycle by adjusting staking incentives.
Carr{\'e} et al.~\cite{carre2024liquid} developed a token trading model to examine the efficiency and security implications of liquid staking within the context of DeFi lending.
Xiong et al.~\cite{xiong2023leverage} investigated the opportunities and risks associated with leveraging staking through liquid staking derivatives within the Ethereum ecosystem.
These studies focus on the analysis of tokenomics, whereas SSR specifically examines the security defects in DeFi staking smart contracts.

\subsection{Smart Contract Defects Detection}
Given the immutable nature of smart contracts, any defects can lead to substantial financial losses once deployed. 
Consequently, numerous studies have focused on detecting smart contract defects. 
These detection methods can be classified into three types: static analysis, dynamic analysis, and formal verification~\cite{qian2022smart}.
Static analysis methods are widely used in contract analysis, including ONENTE~\cite{luu2016making}, ZEUS~\cite{kalra2018zeus}, Securify~\cite{tsankov2018securify}, Slither~\cite{feist2019slither}, Ethainter~\cite{brent2020ethainter}, and Defectchecker~\cite{chen2021defectchecker}.
Dynamic analysis approaches, such as fuzzing, are also commonly used, as demonstrated by Contract-Fuzzer~\cite{jiang2018contractfuzzer}, Mythril~\cite{durieux2020empirical}, FunFuzz~\cite{ye2024funfuzz}, and SmartReco~\cite{zhang2024smartreco}.
Other studies, such as KEVM~\cite{hildenbrandt2018kevm}, Isabelle~\cite{amani2018towards}, and PropertyGPT~\cite{liu2024propertygpt}, employ formal verification methods.


As the complexity of smart contracts increases, new defects continue to emerge. Due to the nature of these new defects, traditional detection tools are often unable to identify them.
Consequently, numerous studies have focused on defining and detecting these new defects. For example, NFTGuard~\cite{yang2023definition}, CrySol~\cite{zhang2024demystifying}, and CDRipper~\cite{lin2024definition} target the detection of defects related to NFTs, cryptographic defects, and centralization defects within smart contracts.

\subsection{Analysis of Logical Defects in Smart Contracts}
Since logical defects in smart contracts are closely tied to the specific logic within the contracts, they cannot be detected by traditional contract analysis tools. 
However, with the recent development of LLMs, the automatic identification of logic in contracts has become possible. As a result, many studies have focused on detecting logical defects within contracts by leveraging this advancement.

GPTScan~\cite{sun2024gptscan} is the first tool to combine GPT with static analysis for detecting logical vulnerabilities in smart contracts. It utilizes GPT as a versatile code comprehension tool and identifies ten specific types of logical vulnerabilities.
S{'o}ley~\cite{soud2024soley} trained a model to detect logical vulnerabilities based on contracts and related code changes from GitHub~\cite{github}.
Liu et al.~\cite{liu2025exploring} collected contracts associated with logical vulnerabilities from 6,165 real audit reports and fine-tuned ChatGPT on this data to improve its ability to detect logical vulnerabilities.
Since these studies do not account for the unique logic of DeFi staking, they are unable to detect the logical defects within it. 
SSR, on the other hand, focuses on the logic of DeFi staking by modeling it specifically and using the resulting model to detect logical defects.

\section{Conclusion}
In this paper, we present an empirical study aimed at identifying logical defects in DeFi staking. 
Through the analysis of 64 security incidents and 144 audit reports, we defined and classified six types of logical defects in DeFi staking. 
Based on this classification, we introduce SSR, a tool specifically designed to identify logical defects in DeFi staking. 
SSR first extracts fundamental information about DeFi staking using large language models (LLMs) and constructs the DeFi staking model. 
It then identifies logical defects by analyzing the model and related semantic features. 
The evaluation results demonstrate that SSR effectively detects logical defects in DeFi staking, achieving an overall precision of 92.31\%, recall of 87.92\%, and an F1-score of 88.85\% on the ground truth dataset. 
Furthermore, SSR revealed that 3,557 (22.24\%) of 15,992 real-world DeFi staking contracts contain at least one logical defect, highlighting the widespread occurrence.

\section*{Acknowledgment}
This work was supported in part by the National Key Research and Development Program of China under Grant 2023YFB2704700, the National Natural Science Foundation of China (No. 62032025, 62302534), the Guangdong Basic and Applied Basic Research Foundation (Grant No. 2025A1515011632), the Major Key Project of Peng Cheng Laboratory under Grant PCL2025AS07.

\bibliographystyle{IEEEtran}
\bibliography{ref}

\vfill

\end{document}